\documentclass[final]{appolb}
\usepackage{graphicx}
\usepackage{amssymb}

\def\muB{\mu_{\mathrm{B}}}
\def\eV{\hbox{ eV}}

\begin{document}
%%%%%%%%%%%%%%%%%%%%%%%%%%%%%%%%%%%%%%%%%%%%%%%%%%%%%%%%%%%%%%%%%%%%%%

\title{Neutrinoless double beta decay mediated by the neutrino magnetic
  moment}

\author{Marek G\'o\'zd\'z, Wies{\l}aw A. Kami\'nski
\address{
Department of Informatics, Maria Curie-Sk{\l}odowska University,
ul. Akademicka 9, 20-033 Lublin, Poland}}

\maketitle

\begin{abstract}
  Neutrinoless double beta decay is a hypothetical nuclear process
  actively developed both on theoretical and experimental grounds. In
  the present paper we extend the idea discussed in
  Ref.~\cite{art25-mg}, where a new channel of this decay has been
  proposed. In this scenario neutrinos not only oscillate inside the
  nucleus but also interact with an external non-uniform magnetic
  field. We assume that the field rotates about the direction of motion
  of the neutrino and show, that for a certain rotation speed the
  half-life of the $0\nu2\beta$ decay can be significantly
  shortened. While the presentation in \cite{art25-mg} was limited to a
  simplified two-neutrino case, in this work we investigate the
  realistic three-neutrino case and perform a detailed numerical study
  of this process.
\end{abstract}

%\keywords{neutrinoless double beta decay, neutrino magnetic moment}

%%%%%%%%%%%%%%%%%%%%%%%%%%%%%%%%%%%%%%%%%%%%%%%%%%%%%%%%%%%%%%%%%%%%%%%%

\section{Introduction}

Neutrinos, although weakly, interact with other particles, and therefore
propagation and oscillation of these particles in vacuum differs from
that in matter. This is known as the Mikheyev -- Smirnov -- Wolfenstein
effect (MSW) \cite{MSW} and has recently been observed by the
Super-Kamiokande Collaboration as an asymmetry in the oscillation rate
between zenith and nadir neutrinos \cite{SKMSW}. This effect is based on
the fact, that the components of `ordinary' matter, i.e., electrons,
protons, and neutrons, interact with electron neutrinos via charged as
well as neutral currents. Muon and tau neutrinos, on the other hand,
cannot interact with the electrons, thus participate in the neutral
current processes only. This results in an asymmetry in the forward
scattering amplitude of different neutrino flavours, effectively
changing the neutrino oscillation parameters. Therefore regular matter
distinguishes between electron and other neutrino flavours.

Weak interactions are not the only factors that may affect neutrino
propagation and oscillations. Despite being electrically neutral,
zeutrinos, according to the Standard Model, should exhibit
electromagnetic properties. In the second order 1-loop process $\nu
\leftrightarrows W^\pm \ell^\mp$ neutrino magnetic moment has been
estimated by Fuijkawa and Shrock to be $3.2 \times 10^{-19} (m_\nu)\muB$
\cite{FujikawaShrock}. For $m_\nu=0.05\eV$ its value reads $1.6 \times
10^{-20} \muB$, $\muB$ being the Bohr magneton \cite{Shrock}. Another
estimations, e.g. by Kayser \cite{Kayser}, provide qualitatively similar
results $\mu\sim 10^{-18}\muB$. This value can be larger in various
scenarios of physics beyond the Standard Model \cite{th-magmom} and for
certain ranges of non-standard parameters it can reach the experimental
limit of roughly $10^{-11}\muB$ \cite{ex-magmom}. Due to the CPT theorem
Majorana neutrinos can have only transition (in the flavour basis)
magnetic moments, while Dirac neutrinos can have also diagonal magnetic
moments. It is important to note that the neutrino-photon effective
interaction vertex may be constructed in such a way, that it violates
the lepton number by two units. This can be realized if one adds
right-handed currents to the Standard Model, in the $R$-parity violating
supersymmetric models and others. In such a situation the transition
magnetic moments change neutrinos into antineutrinos of different
flavour, while the diagonal magnetic moments change neutrinos into
antineutrinos of the same flavour. It has also been pointed out
\cite{B,Giunti2014} that an external non-uniform magnetic field acts
differently on neutrinos and antineutrinos, which is due to different
helicities of these particles. This observation has been used to show
that under special conditions Pontecorvo oscillations \cite{pontecorvo}
$\nu_\alpha \to \bar\nu_\alpha$ are possible \cite{aps93-plb,aps93-prd}.

The neutrinoless double beta decay ($0\nu2\beta$) is a~hypothetical
second order process in which some lepton number violating non-standard
mechanism accounts for the neutrinos not being released. This process is
of the most importance because, if observed, will qualify neutrinos as
Majorana particles, which is the contents of the famous Schechter--Valle
black-box theorem \cite{schechter}. To be more exact, as shown in
\cite{duerr}, the black-box theorem states, that the $0\nu2\beta$ decay
implies a Majorana-like contribution to the neutrino mass matrix, but
does not exclude other contributions, also Dirac-like, which in
principle could be even dominant. Nevertheless, for a pure Dirac
neutrino this decay is strictly forbidden. In this paper we assume
neutrinos to be Majorana particles.

The simplest and most often discussed mechanism of the $0\nu2\beta$
decay is the so-called mass mechanism, in which left-handed Majorana
neutrinos of non-zero mass are produced in the beta vertex as a
negative-helicity state with a small positive-helicity admixture. This
positive-helicity admixture is responsible for the possibility of the
neutrino being absorbed in the second beta vertex. Of course the
intermediate electron neutrino propagates between the beta vertices as a
superposition of three mass eigenstates and therefore the inverse
half-life of the decay depends on the so-called effective neutrino mass
$\langle m \rangle_{0\nu}$. Assuming the exchange of light neutrino and
the same chirality in both beta vertices, the neutrino part of the
process is described by
\begin{equation}
  \sum_{i=1,2,3} U_{ei} \frac{m_i}{p^2-m_i^2} U_{ei}^*
  \approx \frac{1}{p^2} \sum_{i=1,2,3} |U_{ei}|^2 m_i 
  = \frac{1}{p^2} \langle m \rangle_{0\nu},
\label{eq:1}
\end{equation}
where $p$ is the neutrino momentum, $U_{ei}$ are the elements of the
first row of the neutrino mixing matrix, and we have used the
approximation of small, comparing to $p$, neutrino mass. The factor
$1/p^2$ is then absorbed by the nuclear matrix element as a part of the
energy denominator. Other mechanisms involve different intermediate
particles, like the pions, supersymmetric particles, the Majoron, and
others. In this paper we describe another mechanism of the neutrinoless
double beta decay based on the two-step Pontecorvo oscillations. It has
been shown \cite{aps93-plb,aps93-prd} that this mechanism has a
resonant-like behaviour and its application to the solution of the solar
neutrino puzzle has been discussed. In \cite{art25-mg} the very same
mechanism has been discussed in the context of the $0\nu2\beta$ decay
for a simplified two-neutrino case. This paper presents the realistic
three-neutrino case together with a numerical analysis to show, that the
new mechanism may, under proper conditions, significantly shorten the
half-life of the decay in the resonance region.

%%%%%%%%%%%%%%%%%%%%%%%%%%%%%%%%%%%%%%%%%%%%%%%%%%%%%%%%%%%%%%%%%%%%
\section{Neutrinos in nuclear matter}

Our goal is to describe the nuclear process of the neutrinoless double
beta decay. We start therefore with the discussion of the neutrino
behaviour in the nuclear matter. Neutrino interactions and oscillations
inside the nucleus are omitted in the standard approach to the
$0\nu2\beta$ decay \cite{2beta-1,2beta-2}, as it is argued that
neutrinos travel a very short distance between the nucleons. Despite
this fact, the process of flavour oscillations is vital for the proposed
here mechanism.

Neutrinos travelling through matter undergo a phase shift due to their
interactions with electrons, neutrons and protons via neutral and
charged weak currents. A recent discussion of this problem
\cite{kovalenko} takes into account also a specific 4-fermion
interaction of neutrinos and quarks, but we will not consider this
possibility here.

In the most typical case of the MSW effect, matter is electrically
neutral and contains neutrons and an equal amount of electrons and
protons. The charged current Standard Model reactions occur between
charged leptons and the corresponding neutrinos, so typically electrons
and electron neutrinos. Since electrons are absent in the nuclear
medium, this interaction is not present inside the nucleus. The neutral
current contributions coming from the electrons and protons have the
same magnitudes but different signs due to the opposite electric charges
of these particles, resulting in mutual cancelation. In the case of
nuclear matter, however, there are no electrons and the proton
contribution will not be canceled.

We closely follow the textbook approach presented in
\cite{mpal}. Writing the Hamiltonian of neutrinos in vacuum in the mass
basis $(\nu_1,\nu_2,\nu_3)^T$
\begin{equation}
  H = \mathrm{diag}(E_1,E_2,E_3)
\end{equation}
and using the relation for light relativistic particles
$E=\sqrt{p^2+m^2}\approx p + m^2/2p$ we get
\begin{equation}
  H = p + \frac{1}{2p} \ \mathrm{diag}(m_1^2,m_2^2,m_3^2),
\label{eq:tHnu}
\end{equation}
where the symbol $\mathrm{diag}$ represents the diagonal matrix and
$p\equiv|\vec p|$ is the value of the neutrino momentum. The neutrino
mass eigenstates evolve in time according to the Schr\"odinger-like
equation
\begin{equation}
  {\rm i}\frac{{\rm d}}{{\rm d} t}
  \left(\begin{array}{c} \nu_1 \\ \nu_2 \\ \nu_3 \end{array} \right)
  = H
  \left(\begin{array}{c} \nu_1 \\ \nu_2 \\ \nu_3 \end{array} \right).
\end{equation}

The transformation to the flavour basis $(\nu_e,\nu_\mu,\nu_\tau)^T$ is
defined by the unitary Pontecorvo--Maki--Nakagawa--Sakata (PMNS) matrix
$U$ as
\begin{eqnarray}
  H \to U H U^\dagger &=& 
  p + \frac{1}{2p} \left[ U \ \mathrm{diag}(m_1^2,m_2^2,m_3^2) \ 
  U^\dagger \right]
  \nonumber \\
  &=& p + \frac{1}{2p} {\cal M}^2,
  \label{eq:UHU}
\end{eqnarray}
where ${\cal M}^2$ denotes the square of the neutrino mass matrix in the
flavour basis.

The energy levels of the flavour states are corrected in the nuclear
matter by their possible interactions via the neutral currents with
neutrons and protons \cite{mpal},
\begin{equation}
  V_\mathrm{nc} = \sqrt{2} G_\mathrm{F} \sum_{f=\mathrm{n,p}} 
  n_f \left( I_3^{(f)} -2 q^{(f)} \sin^2\theta_\mathrm{W} \right),
\end{equation}
$I_3$ being the third component of the weak isospin, and $q$ is the
electric charge. Explicitly the neutron and proton contributions read
\begin{eqnarray}
  && V^\mathrm{(n)}_\mathrm{nc} = 
  \sqrt{2} G_\mathrm{F} \left(-\frac{1}{2}\right) n_\mathrm{n}, 
  \label{eq:Vn} \\
  && V^\mathrm{(p)}_\mathrm{nc} = \sqrt{2} G_\mathrm{F} 
  \left(\frac{1}{2} - 2\sin^2\theta_\mathrm{W} \right) n_\mathrm{p},
  \label{eq:Vp}
\end{eqnarray}
where $G_\mathrm{F}$ is the Fermi constant, $\theta_\mathrm{W}$ is the
Weinberg mixing angle, and $n_{\mathrm{n,p}}$ are the neutron and proton
number densities. We recall that due to the absence of electrons inside
the nucleus, the charged-current contribution to the energy of electron
neutrino is zero. Therefore all flavour eigenstates are affected by the
presence of nuclear matter in the same way and the contribution in this
specific case takes the form of a constant shift of the neutrino energy
levels. As the neutrino oscillations are sensitive to the differences of
masses squared, this will not affect the oscillation rate.

\section{Neutrinos in an external magnetic field}

As it was already mentioned in the Introduction neutrinos, even within
the Standard Model, possess a non-zero magnetic moment generated in
second-order processes. In the case of Majorana neutrinos this gives a
non-zero probability of the transition between a predominantly
left-handed neutrino $]\nu$ and its right-handed counterpart $\bar\nu$
of different flavour, which is triggered by the effective interaction
with an external photon.

The main mechanism leading to the $0\nu2\beta$ decay is based on the
conversion between different helicities of an electron neutrino. So, if
one combines the interaction via the magnetic moment with the flavour
oscillations, one gets two possible chains which will satisfy the
required conditions $(\alpha=\mu,\tau)$:
\begin{eqnarray}
  && \nu_e \to \bar\nu_\alpha \to \bar\nu_e, 
  \label{eq:chain1} \\
  && \nu_e \to \nu_\alpha \to \bar\nu_e.
  \label{eq:chain2}
\end{eqnarray}
Both of these chains lead to the $0\nu2\beta$ decay, but their
amplitudes cancel each other exactly. It is due to the fact, that the
main part of the amplitudes are the propagators of $\bar\nu_\alpha$ and
$\nu_\alpha$. In normal circumstances these particles have the same
masses, so the propagators are equal, but from the antisymmetricity of
the magnetic moment, $\mu_{e\alpha}=-\mu_{\alpha e}$, follow opposite
signs of the final amplitudes and their cancelation. We have assumed,
however, that there is an external magnetic field with which neutrinos
interact, and behaviour of Majorana neutrinos in these conditions must be
examined.

In the Standard Model neutrinos are massless, which implies that
definite helicity is assigned to a chiral state, so that the left-handed
neutrino has negative helicity, while the right-handed antineutrino has
positive helicity. Such an assignement is in agreement with experiments,
which forces us to define the antineutrino as the CP-conjugate of the
neutrino. Massive left-handed neutrinos, on the other hand, are a
mixture of negative and positive helicity states, where the latter
admixture is proportional to the term $m_\nu/E$, thus being heavily
suppressed for relativistic neutrinos. Therefore for practical reasons
massive neutrinos can be treated as being predominantly in the negative
helicity state and massive antineutrinos being predominantly in the
positive helicity state.

The presence of the magnetic field $B$ has a two-fold effect. Firstly,
transitions between neutrino states of different helicities are possible
via the neutrino magnetic moment. For Majorana neutrinos the magnetic
moment is antisymmetric $\mu_{\alpha\beta} = -\mu_{\beta\alpha}$,
$\alpha, \beta=\{e,\mu,\tau\}$, i.e., only transitions of the form
\begin{equation}
  \nu_\alpha \leftrightarrows \bar\nu_\beta, 
  \qquad \alpha\not=\beta
\end{equation}
are possible. The strength of this interaction has the form
$B\mu_{\alpha\beta}$. The second effect is, that if the magnetic field
changes along the neutrino path, the neutrinos of different helicities
will obtain different corrections to their effective masses. This
follows from the fact that a frame of reference with spin $\vec s$,
which is rotating with the angular velocity $\vec\omega$, gains energy
$(-\vec s \cdot \vec \omega)$. Since left-handed neutrinos ($s=-1/2$)
and their right-handed counterparts ($s=+1/2$) have opposite helicities,
the degeneracy of their energy levels is lifted in the presence of an
external rotating magnetic field. The component of the rotating field
that is parallel to $\vec s$ does not contribute to this effect, so we
denote by $B\equiv|\vec B_\perp|$ the magnitude of the perpendicular
component of the magnetic field, with the angle $\phi=\phi(t)$
indicating the direction of $\vec B_\perp$, and switch to a~reference
frame which rotates with the field. We can write the magnetic field
angular velocity as $\omega = {\rm d}\phi(t)/{\rm d} t \equiv
\dot\phi(t)$. The correction to the energy coming from the field
rotation is $+\dot\phi(t)/2$ for left-handed particles and
$-\dot\phi(t)/2$ for right-handed particles, and the immediate
consequence of the lifting of the degeneracy of Majorana neutrinos of
different helicities masses is, that the amplitudes of the chains
(\ref{eq:chain1})-(\ref{eq:chain2}) do not cancel each other.

%%%%%%%%%%%%%%%%%%%%%%%%%%%%%%%%%%%%%%%%%%%%%%%%%%%%%%%%%%%%%%%%%%%%%%
\subsection{The two-flavour case}

In the two neutrinos case \cite{aps93-plb,aps93-prd,art25-mg} the
results can be presented in an concise analytical form and they exhibit
all the key features of the realistic three-flavour case, which will be
described in the next section. Choosing the basis
\begin{equation}
  (\nu_e,\nu_\mu,\bar\nu_e,\bar\nu_\mu)^T
\end{equation}
the mixing matrix depends on one vacuum mixing angle $\theta$ only:
\begin{equation}
  U = \left(
  \begin{array}{cc} \
    \cos\theta & \sin\theta \\ -\sin\theta & \cos\theta
  \end{array} \right ).
\end{equation}
We neglect here, for simplicity, the possible CP-violating phases, but
will discuss them in the three-neutrino case. The Hamiltonian, diagonal
in the mass basis, becomes non-diagonal in the flavour basis. Taking
into account matter and magnetic field corrections it takes the block
form:
\begin{equation}
  H = \left(
    \begin{array}{cc} \
      H_\nu + \frac{\dot\phi}{2} & [B\mu]_2 \\ 
      -[B\mu]_2 & H_\nu - \frac{\dot\phi}{2}
    \end{array} \right ),
\label{eq:H-2nu}
\end{equation}
where
\begin{equation}
  H_\nu = p + V_\mathrm{nc}^\mathrm{(n)} + V_\mathrm{nc}^\mathrm{(p)} 
  + \frac{1}{2p} {\cal M}^2,
\label{eq:Hnu}
\end{equation}
with
\begin{equation}
{\cal M}^2 = 
  \left(
    \begin{array}{cc}
      m_1^2\cos^2\theta + m_2^2\sin^2\theta & 
      \frac{\Delta m^2}{2}\sin2\theta \\ 
      \frac{\Delta m^2}{2}\sin2\theta &
      m_1^2\sin^2\theta + m_2^2\cos^2\theta
    \end{array} \right ),
\end{equation}
$\Delta m^2 = m_2^2-m_1^2$, $V_\mathrm{nc}^\mathrm{(n)}$ and
$V_\mathrm{nc}^\mathrm{(p)}$ being given by (\ref{eq:Vn}) and
(\ref{eq:Vp}), and
\begin{equation}
  [B\mu]_2 = B \left(\begin{array}{cc} 
      0 & \mu_{e\mu} \\ -\mu_{e\mu} & 0 
    \end{array} \right ).
\end{equation}
Here $\mu_{e\mu}$ is the antisymmetric Majorana neutrino transition
magnetic moment.

The Hamiltonian (\ref{eq:H-2nu}) can be diagonalized and its eigenvalues
take the form
\begin{equation}
  p + V_\mathrm{nc}^\mathrm{(n)} + V_\mathrm{nc}^\mathrm{(p)} 
  + \frac{1}{2p} 
  \left( \begin{array}{l}
      m_1'^2 \\ m_2'^2 \\ \bar m_1'^2 \\ \bar m_2'^2
      \end{array} \right),
\end{equation}
where the overbar indicates masses of the antiparticles. Explicitly the
masses squared are given by
\begin{eqnarray}
  && m_{1,2}'^2 = \frac{1}{2} \left( m_1^2 + m_2^2 \pm
    \sqrt{(4pB\mu_{e\mu})^2 + (2p\dot\phi + \Delta m^2)^2}
  \right), \nonumber \\ 
\label{eq:2n-mass-a} \\
  && \bar m_{1,2}'^2 = \frac{1}{2} \left( m_1^2 + m_2^2 \pm
    \sqrt{(4pB\mu_{e\mu})^2 + (2p\dot\phi - \Delta m^2)^2}
  \right). \nonumber \\
\label{eq:2n-mass-b}
\end{eqnarray}
(There is a typo in Eq.~(10) of Ref. \cite{art25-mg}: the factor 2 is
missing in $2p\dot\phi$.) We notice, that the term $\dot\phi$ lifts the
degeneracy between the mass eigenstates of neutrinos of different
helicities, $m_{1,2}'^2 \not = \bar m_{1,2}'^2$. Also, in the absence of
the magnetic field, $B=0$, $\dot\phi=0$, we arrive at the expected
result $m_{1,2}'^2=m_{1,2}^2=\bar m_{1,2}'^2$.

In the general case, the mixing angles of neutrino mass eigenstates in
vacuum and in matter differ, and the source of this difference lies in
the interaction of electron neutrinos with electrons via the charged
current $V_\mathrm{cc}$. The corrected mixing angle is given by
\begin{eqnarray}
  \tan2\theta' = \frac{\Delta m^2 \sin2\theta}{\Delta m^2 \cos 2\theta 
    - 2 p V_\mathrm{cc}},
\end{eqnarray}
where $V_\mathrm{cc}\sim n_\mathrm{e}$, the electron number density in
matter. In our case, however, $n_\mathrm{e}=0$ and therefore the $U$
matrix remains unchanged. As mentioned earlier, one can take into
account the $\nu\bar\nu q\bar q$ interaction \cite{kovalenko} to refine
the shape of the $U$ matrix. We leave this possibility to be included in
future work.

The nuclear matter effect on neutrino propagation manifests itself as a
constant shift of the mass eigenstates. This shift induces a constant
phase factor which does not affect the oscillation probabilities, but
has an influence on the neutrino propagator.

%%%%%%%%%%%%%%%%%%%%%%%%%%%%%%%%%%%%%%%%%%%%%%%%%%%%%%%%%%%%%%%%%%%%%% 
\subsection{The three-flavour case and the neutrinoless double beta
  decay}

We present the generalization to the realistic three-neutrino case by
expanding the flavour basis of neutrinos and antineutrinos to the form
\begin{equation}
  (\nu_e,\nu_\mu,\nu_\tau,\bar\nu_e,\bar\nu_\mu,\bar\nu_\tau)^T.
\end{equation}
and taking into account the presence of matter and an external
non-constant magnetic field in the way descibed in the previous
section. The three-flavour neutrino Hamiltonian takes the form:
\begin{equation}
  H = \left(
    \begin{array}{cc} \
      H_\nu + \frac{\dot\phi}{2} & [B\mu]_3 \\ 
      -[B\mu]_3 & H_\nu - \frac{\dot\phi}{2}
    \end{array} \right ),
\label{eq:H-3nu}
\end{equation}
where
\begin{equation}
  [B\mu]_3 = B \left( \begin{array}{ccc} 
             0     &  \mu_{e\mu}    & \mu_{e\tau} \\
      -\mu_{e\mu}  &       0        & \mu_{\mu\tau} \\
      -\mu_{e\tau} & -\mu_{\mu\tau} &       0
    \end{array} \right).
\label{eq:Bmu}
\end{equation}
$H_\nu$ is given by Eq.~(\ref{eq:Hnu}) with ${\cal M}^2$ defined in
(\ref{eq:UHU}). We use the standard parametrization for the PMNS matrix,
\begin{equation}
  U = 
\left (
    \begin{array}{ccc}
      c_{12} c_{13} & s_{12} c_{13} & s_{13} e^{-i \delta} \\
      -s_{12} c_{23} - c_{12} s_{23} s_{13}e^{i \delta} & c_{12} c_{23} - s_{12}
      s_{23} s_{13}e^{i \delta} & s_{23} c_{13} \\
      s_{12} s_{23} - c_{12} c_{23} s_{13}e^{i \delta} & -c_{12} s_{23} - s_{12}
      c_{23} s_{13}e^{i \delta} & c_{23} c_{13}
    \end{array}
  \right ) 
  \times \mathrm{diag}(1,e^{i \phi_{2}},e^{i \phi_{3}}),
\label{eq:U}
\end{equation}
where $s_{ij} \equiv \sin\theta_{ij}$, $c_{ij} \equiv \cos\theta_{ij}$,
and $\theta_{ij}$ is the mixing angle between the mass eigenstates $m_i$
and $m_j$. The $\delta$ is the CP violating Dirac phase and $\phi_2$,
$\phi_3$ are CP violating Majorana phases. We will discuss the impact of
these phases in Section \ref{sec:4}.

For the neutrinoless double beta decay the essential neutrino transition
is that of $\nu_e \leftrightarrows \bar\nu_e$. The off-diagonal block of
the Hamiltonian, which describes the mixing of neutrinos of different
helicities in our setup, has the form (\ref{eq:Bmu}) and therefore a
direct transition between $\nu_e$ and $\bar\nu_e$ is forbidden. This
transition, however, is possible, as it was pointed out in
\cite{aps93-plb,aps93-prd}, in a two-step processes given by
(\ref{eq:chain1}) and (\ref{eq:chain2}). In the neutrinoless double beta
decay the chains of transitions have to be realized between two beta
vertices and may be described by a factor, call it $\chi$, which is
related to the half-life of the $0\nu2\beta$ decay:
\begin{equation}
  \left( T_{1/2}^{0\nu} \right)^{-1} = 
  G^{0\nu} |M^{0\nu}|^2 |B\chi|^2,
  \label{eq:T12}
\end{equation}
where $G^{0\nu}$ is the exactly computable phase-space factor and
$M^{0\nu}$ contains the hadronic and electron parts of the
amplitude. The neutrino is interchanged between interaction vertices and
as a virtual particle it is not bound by the mass-shell relation. It is
also transfering momentum between the outgoing electrons and the
external photon, which is attached in the magnetic moment vertex. Its
own 4-momentum has to be integrated out. The full discussion of this
problem, which involves the choice of an appropriate nuclear model to
describe the hadronic part, is beyond the scope of the present
manuscript. In what follows we propose and discuss an approximate form
of the $\chi$ term.

The neutrino part of the amplitude is proportional to two propagators of
the intermediate neutrino and two transition probabilities, i.e., the
magnetic moment and a product of the $U$ matrices. In order to proceed,
we have to make some approximations. Firstly, we neglect the 4-momenta
of the electrons and the photon. The product of two propagators of
neutrinos with a common 4-momentum $q$ reduces then to
\begin{equation}
  \frac{\not q + \bar m_i'}{q^2 - \bar m_i'^2}
  \frac{\not q + m_j'}{q^2 - m_j'^2} \to 
  \frac{q^2 + \bar m_i' m_j'}{(q^2-\bar m_i'^2)(q^2 - m_j'^2)}
\label{eq:props}
\end{equation}
because there are left chiral projectors $P_L$ in each beta vertex, as
the $W$ bosons couple to the left-handed fields only, and $P_L {\not q}
P_L=0$. At this point one usually uses the light neutrino approximation,
see Eq.~(\ref{eq:1}), which is not applicable here, because $m'$ and
$\bar m'$ are functions of $\dot \phi$ and this parameter can be
tuned. We notice, however, that for masses significantly smaller than
the 4-momentum the expression (\ref{eq:props}) is proportional to
$1/q^2$, which is a small number, while it explodes for massess $\bar
m'^2, m'^2 \approx q^2$. This means that in the region close to the
pole, one may retain the dominant terms only and write the approximate
expression for $\chi$ in the form:
\begin{equation}
  \label{eq:chi}
  \chi =
  \sum_{i,j} \sum_{\alpha,\beta} 
   \frac{q^2 + \bar m_i' m_j'}{(q^2-\bar m_i'^2)(q^2-m_j'^2)}
   U_{ei} U_{\alpha i}^* \mu_{\alpha\beta} U_{\beta j}  U_{ej}^*, 
\end{equation}
where $i,j=1,2,3$ number the mass eigenstates and
$\alpha,\beta=e,\mu,\tau$ denote the flavour eigenstates.

We notice, that since the factor $B\mu$ enters the Hamiltonian directly,
it has the unit of mass and therefore $B\chi$ has the unit 1/eV and it
cannot be directly compared with the effective neutrino mass $\langle m
\rangle_{0\nu}$. We discuss the possibility of relating
Eq.~(\ref{eq:chi}) to the expression describing the mass mechanism of
the $0\nu2\beta$ decay in Section \ref{sec:massmech}.

\begin{figure}
  \centering
  \includegraphics[width=0.5\textwidth]{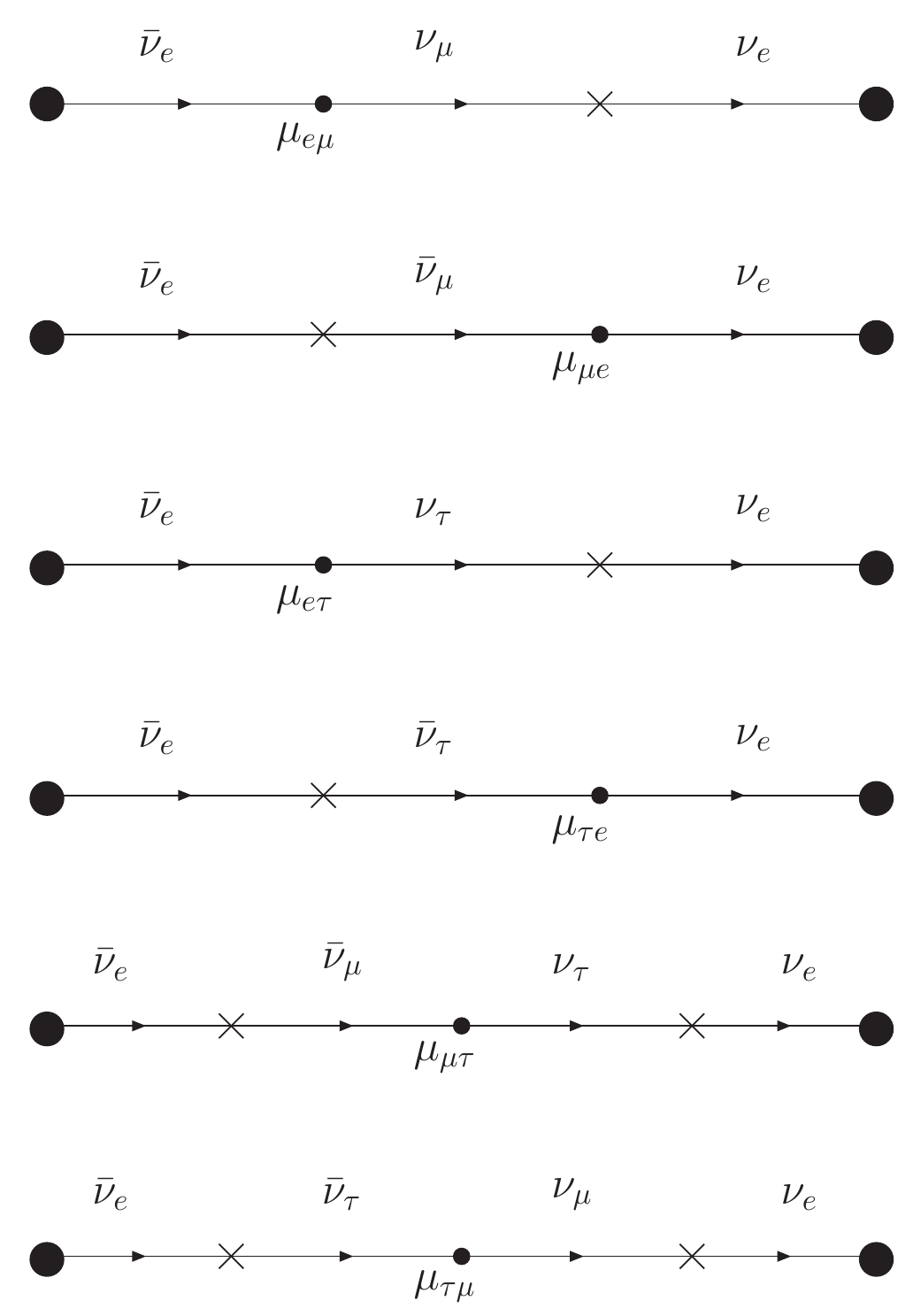}
  \caption{Neutrino transitions that contribute to the neutrinoless
    double beta decay via the magnetic moment. The transitions take
    place in nuclear matter between two beta vertices in the presence of
    a non-stationary magnetic field. The small dots represent the
    magnetic moment insertions while the crosses indicate flavour
    oscillations.}
  \label{fig:1}
\end{figure}

The masses $m'$ and $\bar m'$ are obtained from the diagonalization of
the Hamiltonian (\ref{eq:H-3nu}). All possible choices of the
intermediate flavour states are depicted in Fig.~\ref{fig:1}.

%%%%%%%%%%%%%%%%%%%%%%%%%%%%%%%%%%%%%%%%%%%%%%%%%%%%%%%%%%%%%%%%%%%%
%%%%%%%%%%%%%%%%%%%%%%%%%%%%%%%%%%%%%%%%%%%%%%%%%%%%%%%%%%%%%%%%%%%%

If the masses of the L-handed neutrino and the corresponding R-handed
neutrino are the same, the expression (\ref{eq:chi}) yields zero and
this contribution to the $0\nu2\beta$ vanishes. However, the degeneracy
is removed by different signs of the $\dot \phi$ term in the Hamiltonian
(\ref{eq:H-3nu}). What is more, by having the possibility of changing
this term, we can arrive at the resonance $q^2 \approx m_i'^2$ or $q^2
\approx \bar m_j'^2$ boosting the $\chi$ significantly.

Another interesting observation is that in the case of CP-violation (see
Section \ref{sec:4}) non-zero phases in the matrix $U$,
c.f. (\ref{eq:U}), appear and the expression (\ref{eq:chi}) will not be
zero even if there is degeneracy among the masses. Similar situation
occurs also when there is mixing between the standard model neutrinos
and neutrinos from a fourth generation, in which case the matrix $U$
will no longer be unitary.

%%%%%%%%%%%%%%%%%%%%%%%%%%%%%%%%%%%%%%%%%%%%%%%%%%%%%%%%%%%%%%%%%%%%%%
\section{Numerical analysis \label{sec:4}}

In this section we study the expression (\ref{eq:chi}) numerically. To
start with, we need to compute the corrected neutrino masses by
diagonalizing the Hamiltonian (\ref{eq:H-3nu}). One of the newest
compilations of neutrino oscillation parameters \cite{tortola} gives the
best-fit values:
\begin{eqnarray}
  && \sin^2\theta_{12} = 0.320, \ 
     \sin^2\theta_{13} = 0.026, \ 
     \sin^2\theta_{23} = 0.490, \nonumber \\
  && \Delta m^2_{21} = 7.62 \times 10^{-5} \eV^2, \ 
     \Delta m^2_{31} = 2.53 \times 10^{-3} \eV^2, \nonumber \\ 
\label{eq:numericals}
\end{eqnarray}
where the normal ordering of neutrino masses ($m_1 < m_2 \ll m_3$) is
assumed. At first the CP-violating phases are set to zero. The mass of
the lightest neutrino is arbitrarily set to $m_1=0.05\eV$. We have
estimated the neutrino magnetic moments to be of the order of
$10^{-15}\muB$, the field $B=1$~T, and the neutron and proton number
densities to be $\sim 10^{-31}\eV^3$, which correspond to the $^{76}$Ge
nucleus.

\begin{figure}
  \centering
  \includegraphics[width=0.8\textwidth]{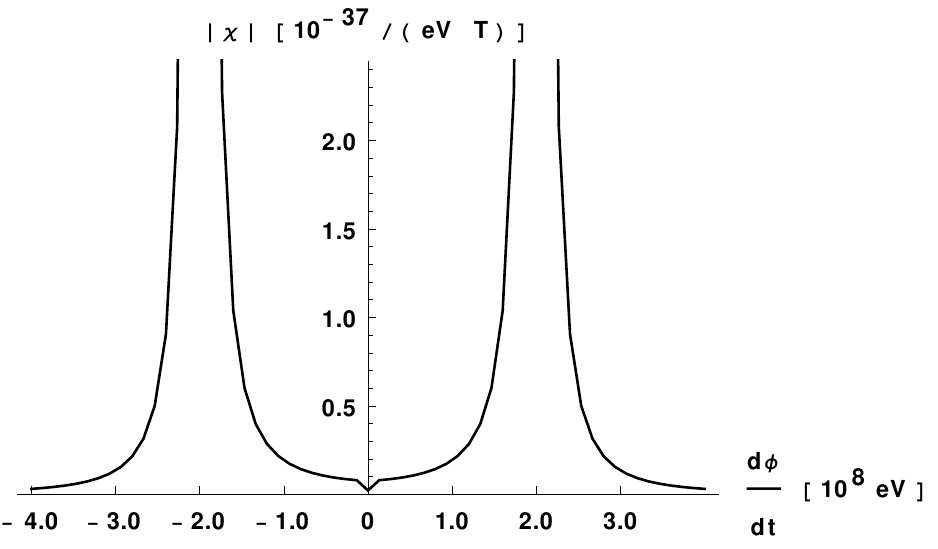}
  \caption{The shape of the function $\chi(\dot\phi)$ for the typical
    neutrino 4-momentum $q=10^8\eV$. A clear resonance peak is visible
    around the value $\dot\phi\approx 2q$.}
  \label{fig:2}
\end{figure}

The average neutrino momentum $\langle p\rangle$ in the $0\nu2\beta$
decay can be assessed from the mean nucleon distance inside the nucleus
and is of the order of 100 MeV \cite{2beta-2}. In what follows we assume
that $q \sim \langle p \rangle = 10^8\eV$. It is this value that must be
compensated by the $\dot\phi$ parameter in order to reach the resonance
region. In the simplified case when due to small mass splitting
($m_1\approx m_2$) one discusses only two neutrino oscillations, it is
possible to assess analytically the resonance condition, which reads
\cite{art25-mg}
\begin{equation}
  \dot\phi \approx 2q.
\end{equation}
In the three-neutrino case the exact formula is much more involved, but
results in a very similar relation. This is illustrated in
Fig.~\ref{fig:2} in which a clear boost in the function $\chi(\dot\phi)$
is visible around the value $\dot\phi\approx 2\times 10^8\eV$. The
figure is symmetric around zero, as the field $B$ may rotate clockwise
or anticlockwise.

%%%%%%%%%%%%%%%%%%%%%%%%%%%%%%%%%%%%%%%%%%%%%%%%%%%%%%%%%%%%%%%%%%%%
%%%%%%%%%%%%%%%%%%%%%%%%%%%%%%%%%%%%%%%%%%%%%%%%%%%%%%%%%%%%%%%%%%%%
%%%%%%%%%%%%%%%%%%%%%%%%%%%%%%%%%%%%%%%%%%%%%%%%%%%%%%%%%%%%%%%%%%%%
\begin{figure}
  \centering
  \includegraphics[width=0.8\textwidth]{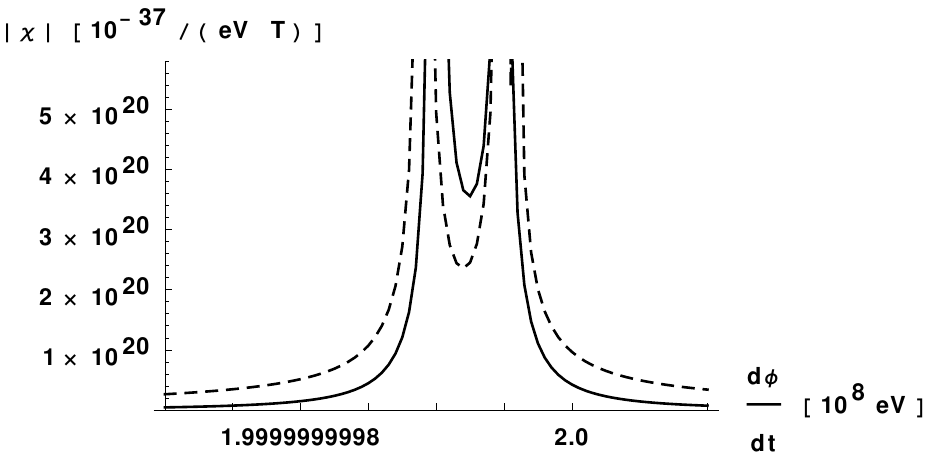}
  \caption{A detailed view of the double structure of the peak from
    Fig.~\ref{fig:2}. The two separate peaks correspond to masses
    $m'_{1,2}$ and $m'_3$. Solid line: CP conserving case. Dashed line:
    the phase $\delta=\pi/2$.}
  \label{fig:3}
\end{figure}
%%%%%%%%%%%%%%%%%%%%%%%%%%%%%%%%%%%%%%%%%%%%%%%%%%%%%%%%%%%%%%%%%%%%

One would expect, that the peaks should appear for each mass eigenstate
separately. This is indeed the case, but in Fig.~\ref{fig:2} the
detailed structure of the peaks is masked by the fact, that the
differences of neutrino masses are much smaller than neutrino momenta.
The structure of the peak is visible in Fig.~\ref{fig:3}, where we have
changed the scale of the horizontal axis. We notice also, that being
close to the pole the value of the $\chi$ parameter increased by 20
orders of magnitude. One sees clearly that what appeared in
Fig.~\ref{fig:2} as a single peak, has a double structure corresponding
to $m'_{1,2}$ and $m'_3$ respectively.

\subsection{CP violating phases}

Up till now we have assumed full CP conservation in the neutrino mixing
matrix and ignored the phases $\delta$, $\phi_2$, and $\phi_3$ in
(\ref{eq:U}). The phases are in principle uncorrelated and independently
take values between 0 and $2\pi$. Global fits from the neutrino
oscillation experiments are rather non-conclusive and the best-fit
values in the $1\sigma$ error range read \cite{tortola}:
\begin{eqnarray}
  & \delta = \left( 0.83^{\ +0.54}_{\ -0.64} \right) \; \pi 
  \qquad & \hbox{\rm (normal hierarchy),} \\
  & \delta = 0.07 \; \pi 
  \qquad & \hbox{\rm (inverted hierarchy),}
\end{eqnarray}
where for the inverted hierarchy case there is no preferred region in
the parameter space. The $2\sigma$ error covers the whole range
$(0,2\pi)$. There are no data for the Majorana phases $\phi_{2,3}$.

\begin{figure}
  \centering
  \includegraphics[width=0.85\textwidth]{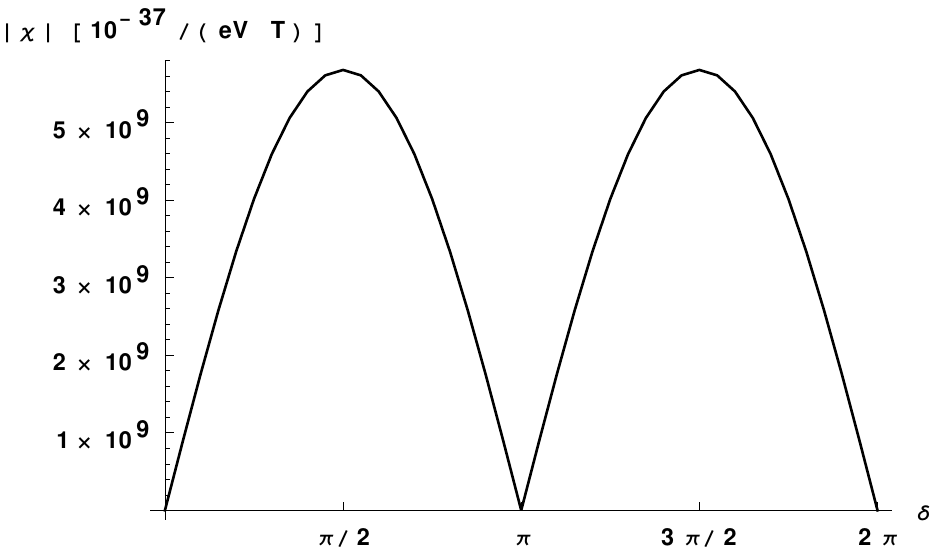}
  \caption{The $|\chi|$ parameter as a function of the Dirac phase
    $\delta$ for $\dot\phi=\langle p\rangle=100$~MeV. The Majorana
    phases are set to zero.}
  \label{fig:4}
\end{figure}

The numerical scale of $\chi$ is dictated by the value of the 4-momentum
$q$, therefore small changes in the $U$ matrix should not change the
$\chi$ behaviour significantly. We have checked, that the Majorana
phases play basically no role, while the Dirac phase introduces a factor
of 6 difference for $\delta=\pi/2,3\pi/2$, see Fig~\ref{fig:4}. In
Fig.~\ref{fig:3} (dashed line) we present the shape of the peak close to
the pole for $\delta=\pi/2$. One sees, that introducing CP violation
does not change the picture qualitatively, while the quantitive changes
are minimal.

%%%%%%%%%%%%%%%%%%%%%%%%%%%%%%%%%%%%%%%%%%%%%%%%%%%%%%%%%%%%%%%%%%%%
%%%%%%%%%%%%%%%%%%%%%%%%%%%%%%%%%%%%%%%%%%%%%%%%%%%%%%%%%%%%%%%%%%%%

%%%%%%%%%%%%%%%%%%%%%%%%%%%%%%%%%%%%%%%%%%%%%%%%%%%%%%%%%%%%%%%%%%%%
\subsection{Neutrino magnetic moment}

The value of the neutrino magnetic moment is bound by observations and
theory. The upper limit comes from dedicated experiments
\cite{ex-magmom} and is given roughly by $10^{-11}\mu_B$. The Standard
Model value is approximately equal to $10^{-19}\mu_B$
\cite{FujikawaShrock,Shrock,Kayser} and everything in between will
originate from some New Physics \cite{th-magmom}.

%%%%%%%%%%%%%%%%%%%%%%%%%%%%%%%%%%%%%%%%%%%%%%%%%%%%%%%%%%%%%%%%%%%%%%
\begin{table}
  \centering
  \caption{\label{tab:1}Approximate numerical values of $|B\chi|$ 
    as a function of $\dot\phi$ and the neutrino magnetic moment, for
    $B=1$~T, $q=100$~MeV, and $\delta=\pi/2$. The resonance points are
    close to the value $\dot\phi=2E$.}
  \begin{tabular}{ccc}
    \hline\hline
    $\dot\phi$& $~~~\mu=10^{-19}\mu_B~~~$ & $~~~\mu=10^{-11}\mu_B~~~$ \\
    \hline
    $1.8 E$ & $5\times 10^{-31}$ & $5\times 10^{-23}$ \\ 
    $1.9 E$ & $1\times 10^{-30}$ & $1\times 10^{-22}$ \\ 
    $2.0 E$ & $2\times 10^{-21}$ & $2\times 10^{-13}$ \\ 
    $2.1 E$ & $1\times 10^{-30}$ & $1\times 10^{-22}$ \\
    \hline\hline
  \end{tabular}
\end{table}
%%%%%%%%%%%%%%%%%%%%%%%%%%%%%%%%%%%%%%%%%%%%%%%%%%%%%%%%%%%%%%%%%%%%%%

The neutrino magnetic moments $\mu$ enter the expression (\ref{eq:chi})
for $|\chi|$ indirectly, by affecting the masses $m'$ and $\bar m'$, and
directly under the sum. We have run a numerical test with the following
parameters:
\begin{eqnarray}
  & q=100 \hbox{\rm\ MeV}, 
  \qquad B=1 \hbox{\rm\ T},
  \qquad \delta=\pi/2, & \nonumber \\ 
  & \qquad \mu_{\alpha\beta} = (10^{-19} - 10^{-11})\ \mu_B. &
\end{eqnarray}
The results for $|B\chi|$ are presented in Tab.~\ref{tab:1}. All
$\mu_{\alpha\beta}$ have been set to a common value $\mu$. The resonance
point is clearly present around $\dot\phi=2E$. One sees also that the
value of $|B\chi|$ scales linearly with $\mu$ which means, that the main
impact comes from the $\mu_{\alpha\beta}$ appearing explicitly in
(\ref{eq:chi}).

\subsection{Comparison with the mass mechanism}
\label{sec:massmech}
 
It is desirable to compare the presented mechanism and the mass
mechanism of the neutrinoless double beta decay. One has to bear in
mind, however, that the nuclear matrix elements (NMEs) should be
calculated for each mechanism separately.

Following \cite{bilenky}, the part of the NME containing neutrino
propagator in the mass mechanism of the $0\nu2\beta$ decay has the form
presented in Eq.~(\ref{eq:1}). The factor $1/p^2$ is successively
integrated out with the help of the assumptions that $p\approx 100$~MeV
and that the average neutrino momentum is much larger than the energy of
the intermediate states (light neutrino approximation). This procedure
allows to factor the expression for the inverse half-life in the
well-known form of a product of the phase-space factor $G^{0\nu}$, the
nuclear matrix element $M'^{0\nu}$, and the effective neutrino mass
$\langle m\rangle_{0\nu}$:
\begin{equation}
  \left( T_{1/2}^{0\nu} \right)^{-1} = 
  G^{0\nu} |M'^{0\nu}|^2 |\langle m\rangle_{0\nu}|^2.
  \label{eq:T12mass}
\end{equation}

Similar factorization is possible in our case leading to
Eq.~(\ref{eq:T12}), but the term $M^{0\nu}$ is not the same as the NME
$M'^{0\nu}$. It follows, that one should compare for a given double beta
emitter the quantities $|M'^{0\nu} \langle m\rangle_{0\nu}|$ and
$|M^{0\nu} B\chi|$. The numerical values of $M^{0\nu}$, however, are not
known and their calculations are beyond the scope of this work.

In order to obtain a crude estimation of the importance of the new
mechanism, we propose the following approximation:
\begin{equation}
  M'^{0\nu} \approx \frac{1}{\langle p\rangle^2} M^{0\nu},
\end{equation}
which allows us to compare the quantities $|B\chi|$ and
$\langle m\rangle_{0\nu}/\langle p\rangle^2$.

%%%%%%%%%%%%%%%%%%%%%%%%%%%%%%%%%%%%%%%%%%%%%%%%%%%%%%%%%%%%%%%%%%%%
\begin{figure}
  \centering
  \includegraphics[width=0.8\textwidth]{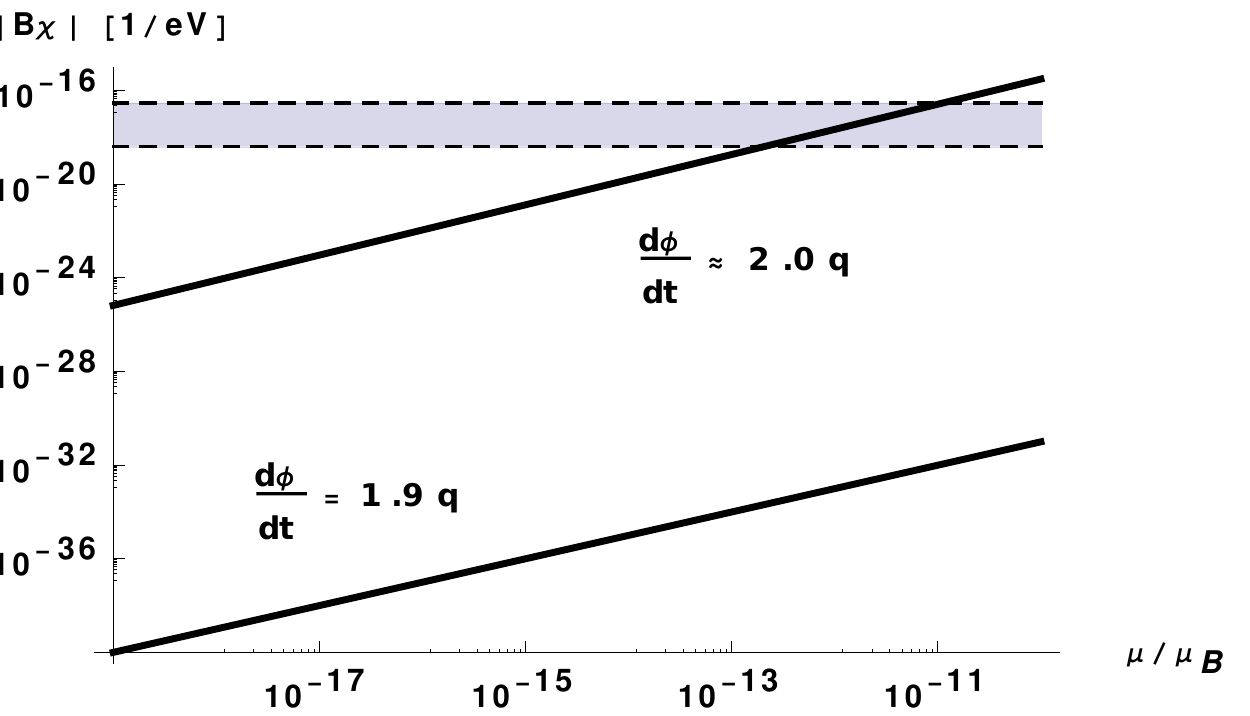}
  \caption{Comparison of the rescaled effective neutrino mass (shaded
    region) with the values of the parameter $|B\chi|$, for $B=1$~T,
    away and close to the resonance. See text for more details.}
  \label{fig:5}
\end{figure}
%%%%%%%%%%%%%%%%%%%%%%%%%%%%%%%%%%%%%%%%%%%%%%%%%%%%%%%%%%%%%%%%%%%%

The neutrino effective mass is given by
\begin{equation}
  \langle m\rangle_{0\nu} = \sum_i |U_{ei}|^2 m_i,
\end{equation}
i.e, it is the sum of mass eigenvalues weighed by the mixing matrix
entries, see Eq.~(\ref{eq:U}). Using Eq.~(\ref{eq:numericals}) and
setting the CP phase $\delta=0$, one obtains the following limiting
values:
\begin{equation}
  4\times 10^{-3} \eV \lesssim
  \langle m \rangle_{0\nu} \lesssim 3\times 10^{-1} \eV,
\end{equation}
which correspond to the mass of the lightest neutrino being 0 and
0.3~eV, respectively. The latter value is motivated by cosmological
observations, which suggest that $\sum m_i \lesssim 1$~eV. We obtain in
this way a rather conservative bound on the possible values of the
effective neutrino mass. It is depicted in Fig.~\ref{fig:5} as the
shaded horizontal band, which has been rescaled by a factor $\langle
p\rangle^{-2}=10^{-16} \eV^{-2}$. The factor $|\chi|$ is a function of
the lightest neutrino mass, the $B$ field, $\dot\phi$, and the neutrino
magnetic moment. In Fig.~\ref{fig:5} we have set $m_1=0.05\eV$,
$q=p=100$~MeV, and other parameters as in Eq.~(\ref{eq:numericals}). On
the horizontal axis of Fig.~\ref{fig:5} the neutrino magnetic moment in
Bohr magnetons is given and the whole range between $10^{-11}\mu_B$ (the
experimental limit) and $10^{-19}\mu_B$ (the Standard Model limit) is
shown. For simplicity we have set all three magnetic moments to the same
value. The straight lines correspond to $\dot\phi=1.9 q$ (lower line),
and $\dot\phi\approx 2.0 q$ (upper line). One can see that close to the
resonance the magnitude of the $|\chi|$ factor is comparable or exceeds
the rescaled effective neutrino mass for the magnetic moment close to
the experimental limit. We notice also that the stronger the field $B$,
the shorter is the expected half-life. Theoretically, in our approximate
approach, an arbitrary low value can be achieved by fine-tuning the
$\dot\phi$ parameter.

As mentioned earlier, a more reliable discussion should include the
recalculation of the NMEs specifically for the presented mechanism. In
the neutrino part, represented here by the $\chi$ parameter, this should
take into account the momenta of the electrons and the photon, a proper
regularization of the propagators, and integration over the neutrino
4-momentum $q$. Finishing this discussion we would like to stress, that
the mass mechanism may be realized spontaneously. Our mechanism,
however, requires certain additional conditions, which have been
described in this paper, and can be realized in a specific environment
only, like in the vicinity of rotating neutron star or magnetar, in
which fast changing magnetic fields of the order of $10^{10}$~T have
been observed.

%%%%%%%%%%%%%%%%%%%%%%%%%%%%%%%%%%%%%%%%%%%%%%%%%%%%%%%%%%%%%%%%%%%%%%

\section{Conclusions}

We have developed a description of a new channel of the neutrinoless
double beta decay, first proposed in our previous work
\cite{art25-mg}. In this scenario electron neutrino emitted in one beta
vertex reacts through its induced magnetic moment with an external
non-uniform magnetic field, which results in a helicity-flip and flavour
change. A subsequent oscillation back to the electron flavour allows for
an absorption of the resulting particle in the second beta
vertex. This process may result in a $0\nu2\beta$ decay.

An interesting feature of this scenario is that the $\chi$ parameter,
which describes the propagating neutrino, depends on the change of the
direction of the external magnetic field, which was represented in our
discussion by the parameter $\dot\phi$. The function $\chi(\dot\phi)$
has a pole and for certain values of the argument, which is roughly
equal to twice the neutrino 4-momentum, becomes large. Since the
half-life of the discussed decay is proportional to $\chi^{-1}$ it
becomes short in the resonance region. This opens, at least
theoretically, the possibility to induce the $0\nu2\beta$ decay by
tuning the external magnetic field.

%%%%%%%%%%%%%%%%%%%%%%%%%%%%%%%%%%%%%%%%%%%%%%%%%%%%%%%%%%%%%%%%%%%% 
\section*{Acknowledgments}

This work has been financed by the Polish National Science Centre
under the decision number DEC-2011/01/B/ST2/05932.

%%%%%%%%%%%%%%%%%%%%%%%%%%%%%%%%%%%%%%%%%%%%%%%%%%%%%%%%%%%%%%%%%%%% 

\end{document}